\documentclass[aps,prl,twocolumn,groupedaddress]{revtex4}
\usepackage{amssymb}

\input{tcilatex}

\begin{document}

\title{Stability of the hydrogen atom of classical electrodynamics }
\author{Jayme De Luca}
\email[author's email address: \ ]{deluca@df.ufscar.br}
\affiliation{Universidade Federal de S\~{a}o Carlos, \\
Departamento de F\'{\i}sica\\
Rodovia Washington Luis, km 235\\
Caixa Postal 676, S\~{a}o Carlos, S\~{a}o Paulo 13565-905}
\date{\today }

\begin{abstract}
We study the stability of the circular orbits of the hydrogen atom in a
version of classical electrodynamics that was historically overlooked. We
introduce the concept of resonant dissipation, i.e. a motion that radiates
the center-of-mass energy while the interparticle distance performs bounded
oscillations about a metastable orbit. The stability mechanism is
established by the existence of a quartic resonant constant generated by the
stiff eigenvalues of the linear stability problem. This constant bounds the
particles together during the radiative recoil. The condition of resonant
dissipation predicts angular momenta for the metastable orbits in reasonable
agreement with the Bohr atom. The principal result is that the emission
lines agree with the predictions of quantum electrodynamics (QED) with 1
percent average error even up to the $40^{th}$ line. Our angular momenta
depend logarithmically on the mass of the heavy body, such that the
deuterium and the muonium atoms have essentially the same angular momenta,
in agreement with QED. Analogously to QED, our stability analysis naturally
uses the eigenvalues of an infinite-dimensional linear operator; the
infinite-dimensionality is brought in by the delay of the electromagnetic
interaction.
\end{abstract}

\pacs{05.45.+b}
\maketitle

According to classical electrodynamics, a quasi-circular orbit of the
hydrogen atom radiates energy.\ We show that this complex Poincar\'{e}%
-invariant two-body motion can damp the center-of-mass energy only, while
the interparticle distance performs bounded oscillations about some circular
orbits, a phenomenon henceforth called resonant dissipation. This dynamics
emits radiation of a sharp frequency while the center-of-mass energy
decreases to compensate for the energy damping, during the life-time of the
emission process, i.e. a collective radiative recoil. The stability
mechanism is established by the existence of a resonant constant that keeps
the interparticle distance bounded during the radiative recoil and
deformation. This resonant normal form exists because of a resonance
involving the stiff modes of the linearized delay equations of motion, i.e.,
the linear normal modes with\ eigenvalues of a very large imaginary
magnitude \cite{StaruszkiewiczPole}. The results of this simple stability
analysis contain no adjustable parameter, satisfy Maxwell's equations and
predict the magnitudes of the Bohr atom\cite{Bohr} with accuracy and
qualitative detail. We calculate every circular emission line of hydrogen
with a $1$ percent error. In this paper we also introduce an electromagnetic
setting slightly different from the most popular form of electrodynamics
with self-interaction and retarded fields. This most popular form is
henceforth called the dissipative-retarded setting (DR). Here we take the
equations of motion derived from Fokker's Lagrangian of the
action-at-a-distance electrodynamics \cite{Schild,StaruszkiewiczPole,Hans},
with advanced and retarded interactions, and we add the Lorentz-Dirac
self-interaction force\cite{Dirac} on each particle. This new
electromagnetic setting is henceforth called the dissipative-Fokker model
(DF). Both the DR and the DF settings satisfy Maxwell's equations everywhere
and define covariant dynamical systems for the electrodynamics of classical
point charges. We experimented with the stability analysis of the several
possibilities within classical electrodynamics, and DF is the most
transparent setting for the resonant dissipation phenomenon introduced here.
Our results advocate for the interest of DF; we predict the Bohr atom with
great precision and qualitative detail. It is popular to use the word
radiation as equivalent to dissipation, because DR electrodynamics is
absolutely dissipative. There is nevertheless a subtle difference from
Poyting's theorem inside DF to Poyting's theorem inside DR; the energy flux
is not necessarily dissipative in DF (because we have advanced and retarded
Li\'{e}nard-Wierchert far-fields). This allows the atom to receive energy
from other atoms of the universe via the effective half/advanced plus
half/retarded far-fields, a thermalization mechanism that is absent in DR.

To understand how our dynamics is a solution of the equations of motion of
classical electrodynamics, one needs to go beyond the simplifications
suggested by the Galilei-invariant Coulomb problem. In quantum
electrodynamics (QED) the semiclassical Bohr orbits of hydrogen correspond
to excited quantum states that decay to the ground state in a life-time of
about $10^{6}$ turns\cite{Bohr}. The analog of this decay process is
troublesome in DR and subtle in DF. Disregarding the singular coupling to
the center-of-mass coordinate is a naive use of the equations of motion of
DR, which predicts that the radiated energy comes out of the electrostatic
potential energy. In this process the interparticle distance decreases while
the rotation frequency increases by a factor of two during the life-time,
making the emission of a sharp line impossible. It is very instructive to
estimate this energy dissipation along hypothetical circular orbits directly
from the Lorentz-Dirac self-interaction force\cite{Dirac}. The complexity
disregarded by this simplification stems from three main sources: \ (i) The
group theoretical complexity of Poincar\'{e}-invariant two-body dynamics;
the two no-interaction theorems\cite{nointeract} limit Hamiltonian and local
Lagrangian descriptions of two-body dynamics to straight-line free motions
only. These negative theorems are sound warning that the Galilei-invariant
Coulombian dynamics, with trivial and separable center-of-mass motion, is a
bad approximation to relativistic two-body dynamics. (ii) The delay
equations of motion bring in an infinite-dimensional dynamical system and
needs an initial function as the initial condition; addition of the
non-runaway condition leads to a well-posed mathematical problem \cite%
{Driver}. The stability analysis of such dynamics reveals stiff eigenvalues
with an arbitrarily large imaginary part, as long as integer multiples of $%
\pi $ (an arbitrary integer number appears naturally) \cite%
{StaruszkiewiczPole}. (iii) The generic solution to a stiff differential
equation is not slowly varying everywhere and typically it has jumps; see
for example the equations of Lienard type in Ref.\cite{Grasman}. Another
example is the small parameter multiplying the highest derivative of the
Lorentz-Dirac self-interaction force\cite{Dirac}. After Dirac's derivation
of a finite force of self-interaction within classical electrodynamics\cite%
{Dirac}, Eliezer considered a model two-body motion with Coulomb interaction
plus self-interaction\cite{Eliezer}. A simple physical consequence
transpired, that with self-interaction the electron cannot fall into the
proton even along unidimensional collision orbits. This surprising result is
a warning that self-interaction is a singular perturbation over the
Coulombian dynamics and also suggests that colinear orbits are the natural
attractors of the dissipative dynamics (a ground state with zero angular
momentum). Along such orbits the heavy particle (the proton) moves in a
non-Coulombian way and the self-interaction provides a repulsive mechanism
that avoids a collision. Disregard of the stiff nature of the equations of
classical electrodynamics has produced several conundrums, and the present
work is an attempt to recognize the correct qualitative behavior of the
solutions of classical electrodynamics.

In Ref. \cite{PRL} we studied the classical helium atom in the DR setting.
Because the very special circular orbits of helium do not radiate in dipole
inside DR, the physics of the emission of a sharp line is easy to understand
even in the DR setting\cite{PRL}. There is already an interesting resonant
structure for these circular obits of helium within the Darwin approximation
(a low-velocity approximation to Fokker's Lagrangian) \cite{PRL,normalDeluca}%
\ \ The predictions of the Darwin approximation were tested numerically in
Ref. \cite{discrete}, finding non-ionizing states in the atomic magnitude,
as predicted in Ref. \cite{PRL}. The generalization to the circular orbits
of the hydrogen atom in the DR setting is troublesome; circular orbits of
hydrogen are stable within the Darwin approximation and radiate in dipole,
such that a sharp emission line seems like a conundrum for classical
electrodynamics in the DR setting. In this work we come to grips with the
physics of the hydrogen atom by changing to DF \emph{and} by understanding
that the correct description comes only after the inclusion of the nuclear
motion (i.e. a two-body system with delay and dissipation). The resonant
dissipation radiates mostly the center-of-mass energy while a slow
deformation of circularity and a collective recoil take place. After a
possibly long life-time, this resonant dissipation produces enough
deformation to destabilize the orbit, i.e. only metastability is possible
(this is also what QED says about circular orbits).

Our perturbation theory starts from the unperturbed circular orbit of the
isolated electromagnetic two-body problem derived from Fokker's Lagrangian,
as explained in Refs.\cite{Schild,StaruszkiewiczPole,Hans}. A motion of the
particles in concentric circles with a constant angular speed and along a
diameter is a solution of these equations of motion. This is because the
symmetric contributions from future and past generate a resulting force that
is normal to the velocity of each particle. We henceforth call this orbit
the circular orbit\cite{Schild, StaruszkiewiczPole,Hans}. The quantities for
this uniform circular motion are given in \cite{Schild}. Our electron has
mass $m_{1}$ and travels a circle of radius $r_{1}$ while the proton has
mass $m_{2}$ and travels a circle of radius $r_{2}$. We henceforth use a
unit system where the electronic charge is $e=-1$ and the speed of light is $%
c=1$. The circular orbit is defined by the retardation angle $\theta $ that
one particle turns while the light emanating from the other particle reaches
it (one light-cone distance away). The frequency $\Omega $ of angular
rotation is given, to lowest order in $\theta $, by Kepler's law%
\begin{equation}
\Omega =\mu \theta ^{3}+...  \label{orbital}
\end{equation}%
where $\mu \equiv m_{1}m_{2}/(m_{1}+m_{2})$ is the reduced mass. For
shallow-energy atomic orbits, the angular momentum $l_{z}$ is given to first
order by $l_{z}^{{}}=\theta ^{-1}$ \cite{Schild}, which is of the order of
the inverse of the fine structure constant, $\alpha ^{-1}=$ $137.036$. To
simplify the stability analysis, we transform the particle coordinates from
Cartesian to gyroscopic coordinates

\begin{eqnarray}
x_{k}+iy_{k} &\equiv &\exp (i\Omega t)[d_{k}+\eta _{k}],  \label{gyroscopic}
\\
x_{k}-iy_{k} &\equiv &\exp (-i\Omega t)[d_{k}+\xi _{k}],  \nonumber
\end{eqnarray}%
where $\eta _{k}$ and $\xi _{k}$ are complex numbers defining the
perturbation of the circularity. The coordinates of the electron are defined
by Eq. (\ref{gyroscopic}) with $k=1$ and $d_{1}=r_{1}$ while the coordinates
of the proton are defined by Eq. (\ref{gyroscopic}) with $k=2$ and $%
d_{2}=-r_{2}$ (at the same time in the inertial frame the particles are in
diametrically opposed positions along the unperturbed orbit \cite{Schild}).
We also need here the stability along the perpendicular $z$ direction, a
linear problem that is uncoupled from the $xy$ stability and easier to work
out. The linear stability analysis proceeds like in Ref. \cite%
{StaruszkiewiczPole}, by substituting the circular orbit plus a
perturbation, Eq. (\ref{gyroscopic}), into Fokker's Lagrangian, expanding
this Lagrangian to quadratic order and taking the Euler-Lagrange equations
to linear order\cite{to be published}. Because we have a delay system at
hand, the normal mode condition (the characteristic equation) has infinitely
many roots. This characteristic equation involves hyperbolic functions that
can become arbitrarily large and are insensitive to an arbitrarily large
imaginary part of the eigenvalue \cite{StaruszkiewiczPole}. A standard
technique of delay analysis is to keep only the largest powers of the
eigenvalue plus the hyperbolic terms in the characteristic equation\cite%
{Bellman}. This large-imaginary-part stiff limit is performed in Ref. \cite%
{StaruszkiewiczPole} for the equal-mass two-body system (see Eq. (15) of
Ref. \cite{StaruszkiewiczPole}). Here we generalize this
large-imaginary-part limit to the arbitrary-mass case and we computed the
characteristic equation with a symbolic algebra software. We define the
normal mode eigenvalue by $\lambda \Omega /\theta $,  i.e. every coordinate
perturbation oscillates in time as $\exp (\lambda \Omega t/\theta )$ ( $%
\lambda $ is an arbitrary complex number). The limiting form of the
characteristic equation for the isolated different-mass case is

\begin{equation}
(\frac{\mu \theta ^{4}}{M})\cosh ^{2}(\lambda )=1,  \label{Istar}
\end{equation}%
where $\mu $ is the reduced mass and $M\equiv m_{1}+m_{2}$. Both the planar
and the perpendicular linearized equations have the same limiting
characteristic Eq. (\ref{Istar}) along circular orbits. For hydrogen $(\mu
/M)$ is a small factor of about $(1/1824)$. It is important to understand
the structure of the roots of Eq. (\ref{Istar}) in the complex $\lambda $
plane, specially for $\theta $ of the order of the fine structure constant.
It is easy to see that the very small parameter $\frac{\mu \theta ^{4}}{4M}%
\sim 10^{-13}$ multiplying the hyperbolic cosine on the left-hand side of
Eq.(\ref{Istar}) determines that $\sigma \equiv |\func{Re}(\lambda )|$ $%
\simeq \ln (\sqrt{\frac{4M}{\mu \theta ^{4}}})$. For the first $13$ excited
states of hydrogen this $\sigma $ is in the interval $14.2<|\sigma |<18.2$.
The imaginary part of $\lambda $ can be an arbitrarily large multiple of $%
\pi $, such that the general solution to Eq. (\ref{Istar}) is 
\begin{equation}
\lambda =\pm (\sigma +i\pi q),  \label{unperastar}
\end{equation}%
where $q$ is an arbitrary integer. \ The plus or minus sign of Eq. (\ref%
{unperastar}) is related to the time-reversibility of the isolated two-body
system, a symmetry that is broken by radiation.

Next we include the dissipation of the DF electrodynamics, i.e. the
Lorentz-Dirac self-interaction, a calculation performed by adding the
self-interaction force to the Lagrangian equations of motion of the isolated
system. Here we give only the limiting form of the characteristic equation%
\begin{equation}
(1-\frac{g}{\lambda ^{2}})(\frac{\mu \theta ^{4}}{M})\cosh ^{2}(\lambda )=1-%
\frac{2}{3}\theta ^{2}\lambda +\frac{1}{\lambda }(\frac{\mu \theta ^{4}}{M}%
)\sinh (2\lambda )+...  \label{ZXY}
\end{equation}%
where $g=-7$ for the planar stability and $g=1$ for the perpendicular
stability. We give the terms up to $O(1/\lambda ^{2})$\ \ because these are
essential for the mathematical phenomenon presented. The coefficient $g$ of $%
\ $the term of order $1/\lambda ^{2}$ is the only\ difference between the
planar and the perpendicular characteristic equations up to $O(1/\lambda
^{2})$. Further terms of size $1/\lambda ^{3}$ and $\theta ^{4}\lambda ^{2}$
produce only small quantitative changes for $\sigma $ in the atomic range
and will be given elsewhere. The linear term on the right-hand side of Eq. (%
\ref{ZXY}) with the $2/3$ coefficient is due to the self-interaction force.
This dissipative term breaks the time-reversal symmetry of Eq. (\ref{Istar})
and the roots of Eq. (\ref{ZXY}) no longer come in plus or minus pairs. In
the following we implement the necessary condition for a convex analytic
function of the interparticle distance to exist.

In the same way already used in Ref.\cite{PRL}, this condition should
involve the perpendicular direction $z$ as well. This is because if the atom
is to recoil like a rigid body while it radiates, one expects the orbital
plane to nutate and precess, typical gyroscopic motions of a rigid body. To
introduce the main idea simply, we assume that the non-runaway condition
restricts the physics to a finite number of linear modes. We take a
perpendicular normal mode and a planar normal mode of Eq. (\ref{ZXY}), of
eigenvalues $\lambda _{z}$ and $\lambda _{xy}$ respectively. Suppose we can
find these modes such that $\lambda _{z}+\lambda _{xy}+\lambda _{z}^{\ast
}+\lambda _{xy}^{\ast }=0$, which is the necessary condition for a quartic
normal form. i.e.%
\begin{equation}
\func{Re}(\lambda _{z}+\lambda _{xy})=0.  \label{necessary}
\end{equation}%
The coordinate of the planar normal mode is a linear combination of the four 
$\eta \xi $ gyroscopic coordinates: $u\equiv a_{1k}\eta _{k}+b_{1k}\xi _{k}$
,while the coordinate of the perpendicular $z$ normal mode is $Z\equiv
b_{1}z_{1}+b_{2}z_{2}$. Because Fokker's Lagrangian is real, $\lambda
_{z}^{\ast }$ and $\lambda _{xy}^{\ast }$ are also solutions to Eq. (\ref%
{ZXY}) with complex conjugate normal mode coordinates. Using these normal
mode coordinates and Eq. (\ref{necessary}), one can show that the following
quartic form is a constant of motion up to higher order terms\cite%
{normalDeluca}: 
\begin{equation}
C\equiv |u|^{2}|Z|^{2}+...  \label{normalform}
\end{equation}%
Notice from Eq. (\ref{gyroscopic}) that $|u|^{2}$ is a quadratic function of
the planar coordinates, such that the resonant constant of Eq. (\ref%
{normalform}) limits the excursion along the $z$ direction times the planar
distance to the circular orbit, essentially locking the dynamics to the
circular orbit. This necessary condition and the continuation of the leading
term (\ref{normalform}) to an asymptotic series is discussed in Ref. \cite%
{normalDeluca}.\ The lesson of the above construction is that if the
interparticle distance is to be bound by a resonant constant while the
dissipation goes on, then we must be able to find a pair of roots to Eq. (%
\ref{ZXY}) such that Eq. (\ref{necessary}) holds. This implies that Eq. (\ref%
{ZXY}) has a pair of almost-symmetric roots of the form%
\begin{eqnarray}
\lambda _{xy} &\equiv &(\sigma +\pi qi+i\epsilon _{1}),  \label{pair} \\
\lambda _{z} &\equiv &-(\sigma +\pi qi+i\epsilon _{2}),  \nonumber
\end{eqnarray}%
where $\epsilon _{1}$ and $\epsilon _{2}$ can be assumed real if $\lambda
_{z}+\lambda _{xy}$ is purely imaginary by absorbing the real part of the
perturbation in the definition of $\sigma $. The above-defined
root-searching problem is well posed and for each integer $q$ conditions (%
\ref{ZXY}) and (\ref{pair}) determine $\theta $ as a function of $q$ , i.e., 
$\theta $ must be quantized! An asymptotic solution to condition (\ref{pair}%
) can be obtained by expanding Eqs. (\ref{ZXY}) up to quadratic order in $%
\epsilon _{1}$ and $\epsilon _{2}$ while treating $\sigma $ as an
approximate constant. This approximation determines the following discrete
values for $\theta $

\begin{equation}
\theta ^{2}=\frac{6(\pi ^{2}q^{2}-\sigma ^{2})}{\sigma (\pi ^{2}q^{2}+\sigma
^{2})^{2}},  \label{asintheta}
\end{equation}%
and 
\begin{equation}
(\epsilon _{1}-\epsilon _{2})=\frac{4\pi q(3\sigma ^{2}-\pi ^{2}q^{2})}{%
\sigma (\sigma ^{2}+\pi ^{2}q^{2})^{2}}.  \label{e1e2}
\end{equation}%
According to QED, circular Bohr orbits have maximal angular momentum and a
radiative selection rule ( $\Delta l=\pm \hbar $) restricts the decay from
level $k+1$ to level $k$ only, i.e. circular orbits emit the first line of
each spectroscopic series (Lyman, Balmer, Ritz-Paschen, Brackett, etc...),
henceforth called the QED circular line. We have solved Eqs. (\ref{pair})
and (\ref{ZXY}) with a Newton method in the complex $\lambda $ plane. Every
angular momentum $l_{z}=\theta ^{-1}$ determined by Eq. (\ref{necessary})
has a value in the correct atomic magnitude, but the subset of Table 1 has
frequencies $w_{DF}$ surprisingly close to the QED lines. These lines are
when the integer $q$ is approximately equal to an integer multiple of the
integer part of $2\sigma $. The reason for such selection rule is that from
the resonances satisfying the necessary condition, only some have $|u|^{2}$
depending on the translation-invariant quantities $(\xi _{1}-\xi _{2})$ and $%
(\eta _{1}-\eta _{2})$ to allow a recoiling translation\cite{to be published}%
. In our description the emission mechanism is at a frequency equal to the
orbital frequency $\Omega $ corrected by the frequency of the complex
amplitude $uZ$ defined above Eq. (\ref{normalform}), as we explain below.
The numerically calculated angular momenta $l_{z}=\theta ^{-1}$ for this
select subset are given in Table 1, along with the orbital frequency in
atomic units $(137^{3}\Omega )/\mu =(137\theta )^{3}$, the QED first
frequency of the series in atomic units $w_{QED}\equiv \frac{1}{2}(\frac{1}{%
k^{2}}-\frac{1}{(k+1)^{2}})$, and the frequency predicted by the dissipative
Fokker model $w_{DF}\equiv (137\theta )^{3}+137^{3}\theta ^{2}(\epsilon
_{1}-\epsilon _{2})$. We list only the first $13$ lines, which are the
experimentally observable, but we tested the agreement of the numerical
calculations of the Newton method with up to the $40^{th}$ circular line
predicted by QED. Beyond that, the asymptotic formula (\ref{asintheta})
shows that the agreement is essentially for any integer $k$ because
substitution of $q=[2\sigma ]k$ into Eq. (\ref{asintheta}) yields 
\begin{equation}
\theta ^{-1}=\sqrt{\frac{2\pi ^{2}}{3}}\sigma ^{3/2}k\sim 137.9k,
\label{hbar}
\end{equation}%
to be compared with the $137.036$ of QED. The agreement for any integer $k$
suggests that Eq. (\ref{ZXY}) is equivalent to Schroedinger's equation (both
are eigenvalue problems for an infinite-dimensional linear operator, and
linear operators with the same spectrum are equivalent).

\bigskip

\begin{tabular}{|l|l|l|l|}
\hline
$l_{z}=\theta ^{-1}$ & $(137\theta )^{3}$ & $w_{QED}$ & $w_{DF}$ \\ \hline
161.21 & 6.137$\times $10$^{-1}$ & 3.750$\times $10$^{-1}$ & 3.688$\times $10%
$^{-1}$ \\ \hline
282.88 & 1.136$\times $10$^{-1}$ & 6.944$\times $10$^{-2}$ & 6.806$\times $10%
$^{-2}$ \\ \hline
397.40 & 4.097$\times $10$^{-2}$ & 2.430$\times $10$^{-2}$ & 2.470$\times $10%
$^{-2}$ \\ \hline
519.59 & 1.833$\times $10$^{-2}$ & 1.125$\times $10$^{-2}$ & 1.113$\times $10%
$^{-2}$ \\ \hline
637.79 & 9.911$\times $10$^{-3}$ & 6.111$\times $10$^{-3}$ & 6.054$\times $10%
$^{-3}$ \\ \hline
751.48 & 6.059$\times $10$^{-3}$ & 3.685$\times $10$^{-3}$ & 3.719$\times $10%
$^{-3}$ \\ \hline
871.84 & 3.880$\times $10$^{-3}$ & 2.391$\times $10$^{-3}$ & 2.392$\times $10%
$^{-3}$ \\ \hline
987.25 & 2.672$\times $10$^{-3}$ & 1.640$\times $10$^{-3}$ & 1.654$\times $10%
$^{-3}$ \\ \hline
1109.18 & 1.188$\times $10$^{-3}$ & 1.173$\times $10$^{-3}$ & 1.170$\times $%
10$^{-3}$ \\ \hline
1225.92 & 1.139$\times $10$^{-3}$ & 8.678$\times $10$^{-4}$ & 8.694$\times $%
10$^{-4}$ \\ \hline
1343.20 & 1.061$\times $10$^{-3}$ & 6.600$\times $10$^{-4}$ & 6.628$\times $%
10$^{-4}$ \\ \hline
1466.89 & 8.146$\times $10$^{-4}$ & 5.136$\times $10$^{-4}$ & 5.102$\times $%
10$^{-4}$ \\ \hline
1585.51 & 6.455$\times $10$^{-4}$ & 4.076$\times $10$^{-4}$ & 4.052$\times $%
10$^{-4}$ \\ \hline
\end{tabular}

\bigskip

Table 1: \ Numerically calculated angular momenta $l_{z}=\theta ^{-1}$ in
units of $e^{2}/c$, orbital frequencies in atomic units $(137\theta )^{3}$,
circular lines of QED in atomic units $w_{QED}\equiv \frac{1}{2}(\frac{1}{%
k^{2}}-\frac{1}{(k+1)^{2}})$ and the emission frequencies of the DF model in
atomic units $w_{DF}\equiv (137\theta )^{3}+137^{3}\theta ^{2}(\epsilon
_{1}-\epsilon _{2})$ .

Last we calculate the frequency of the emission line. The treatment is a
consequence of the above defined DF version of electrodynamics and differs
from the usual electromagnetic model (DR) in non-trivial ways. We have
consistently used the relativistic action-at-a-distance electrodynamics as a
many-body electromagnetic theory, as first suggested in Ref. \cite{Fey-Whee}%
. This no-field theory is based on a parametrization-invariant action
involving two-body interactions only, without the mediation by fields. The
isolated electromagnetic two-body problem, away from the other charges of
the universe, is a time-reversible dynamical system defined by Fokker's
action 
\begin{eqnarray}
S_{F} &=&-\int m_{1}ds_{1}-\int m_{2}ds_{2}  \label{Fokker} \\
&&+\int \int \delta (||x_{1}-x_{2}||^{2})\dot{x}_{1}\cdot \dot{x}%
_{2}ds_{1}ds_{2},  \nonumber
\end{eqnarray}%
where $x_{i}(s_{i})$ represents the four-position of particle $i=1,2$
parametrized by its arc-length $s_{i}\,$, double bars stand for the
four-vector modulus $||x_{1}-x_{2}||^{2}\equiv (x_{1}-x_{2})\cdot
(x_{1}-x_{2})$, and the dot indicates the Minkowski scalar product of
four-vectors with the metric tensor $g_{\mu \nu }$\ ($%
g_{00}=1,g_{11}=g_{22}=g_{33}=-1$). The Euler-Lagrange equations of
Lagrangian (\ref{Fokker}) were already used to derive our Eq. (\ref{Istar}).
Surprisingly, the solution to the unusual equations of motion of Lagrangian (%
\ref{Fokker}) are still determined by initial position and velocity, as
proved in Ref.(\cite{Driver}) for the non-runaway solutions. Therefore we
are dealing with a perfectly well-posed and causal system dressed in an
unfamiliar form, and we invite the reader to consult Ref. \cite{Driver}. In
the many-body action-at-a-distance electrodynamics one needs a prescription
to describe the incoherent interaction with the universe, because including
every single particle of the universe in a pairwise many-body Lagrangian
like Eq. (\ref{Fokker}) is impractical (even though it is precisely what we
want to describe physically). This prescription for the dissipative
interaction with the universe must be a covariant prescription that
satisfies Maxwell's equations. For example, the Wheeler-Feynman prescription
of Ref. \cite{Fey-Whee} leads to the usual electromagnetic DR model. The
prescription of the DF model is to add the Lorentz-Dirac self-interaction
directly to the Euler-Lagrange equations of action (\ref{Fokker}).

In the DF model the interaction with a distant particle involves
half/retarded plus half/advanced Li\'{e}nard-Wierchert interactions, as
derived from a pairwise Lagrangian of type (\ref{Fokker}). We shall show
that this does not preclude radiation of energy. Once DF satisfies Maxwell's
equations everywhere, Poynting's theorem is valid for any true orbit that
solves the stiff equations of motion. Thinking of Poynting's theorem as
detached from the complexity of the equations of motion might suggest a use
of \ this theorem to calculate the radiation of hypothetical orbits that are
not even solution to the equations of motion. For example, one can show that
with half/retarded plus half/advanced fields a perfectly circular orbit does
not radiate energy. But again, it is just as easy to show that a perfectly
circular orbit is not a solution of the equations of motion with the
self-interaction terms. Stiffness can prevent the dynamics from staying even
for a single turn in the neighborhood of an orbit that would be a solution
to the equation without the stiff terms \cite{Grasman}. A familiar use of
the retarded-only radiation fields of DR is to calculate the incoherent
dissipation of a circular current by applying Poynting's theorem to the
radiation fields of a hypothetical circular orbit. It is a trivial exercise
to show that this produces the same result of integrating the Lorentz-Dirac
self-interaction times the velocity along that hypothetical circular orbit.
Even though this approximation is fine for macroscopic currents restricted
by a wire and forced by a battery, the stiff delay equations of
electrodynamics can jump wildly given unstable conditions. Therefore, one
should see first if the solution exists and if it is stable in the model and
only then calculate the radiation for that precise stable orbit. The
perturbations to circularity defined by Eq. (\ref{gyroscopic}) are necessary
to stabilize the orbit, and as we show below, are essential as well to
determine the radiated energy. The self-interaction force is a very small
singular dissipative term along the unperturbed circular orbit, and the
dissipation calculated over the perturbed orbit is completely changed by the
gyroscopic perturbations, even if the final orbit stays near the circular
orbit. After proper account of these perturbations, our metastable orbit
radiates even with the half-retarded plus half-advanced fields.

The half/retarded plus half/advanced radiation magnetic field of the
electron in the DF model (the far-magnetic field) is\cite{Rohrlich} 
\begin{equation}
B_{rad}=\frac{(a_{-}\times \hat{n}_{-})}{2(1-\hat{n}_{-}\cdot v_{-})^{2}r}-%
\frac{(a_{+}\times \hat{n}_{+})}{2(1-\hat{n}_{+}\cdot v_{+})^{2}r},
\label{Bfar}
\end{equation}%
where $v$ and $a$ are the electronic velocity and acceleration, $\hat{n}$ is
a unit vector from the electron to the observation point, underscore minus
indicates evaluation on the retarded light-cone and underscore plus
indicates evaluation on the advanced light-cone. These two light-cones are
defined by $t_{\pm }=t\pm (r-\hat{n}_{\pm }\cdot x)$ where $x$ is the vector
position of the electron measured from the center of the circular orbit.
Once $(t_{+}-t_{-})\simeq $ $2r$, whenever $2r$ is an integer multiple of
the distance travelled by light in a circular period (one light-period), the
first term of expansion \ (\ref{Bfar}) along the precise circular orbit
vanishes and the next term is the important one; a quadratic function of the
orbital quantities. The same cancellation happens for the far-electric field
whenever $2r$ is an odd-integer multiple of half the light-period, but for
us here the far-magnetic field is enough. The next term of the magnetic
expansion about the circular orbit is%
\begin{equation}
B_{rad}^{(1)}=\frac{2(\hat{n}\cdot v)(a\times \hat{n})+(\hat{n}\cdot x)(\dot{%
a}\times \hat{n})}{r}.  \label{quadraticRad}
\end{equation}%
We can estimate (\ref{quadraticRad}) by noticing that along the $\hat{n}%
_{\pm }=\hat{x}$ direction of the unperturbed plane this quadratic term
contains a product of the $z$ perturbed coordinate times the $x$ perturbed
coordinate. Those perturbations are described by the $u$ and $Z$ nonrunaway
normal modes already used and explained above Eq. (\ref{normalform}). Using
the normal mode conditions $\theta \dot{u}=\Omega \lambda _{xy}u$ and $%
\theta \dot{Z}=\Omega \lambda _{z}Z$ together with Eq.(\ref{pair}) one can
show that the quadratic form $uZ$ is a complex amplitude that oscillates
harmonically with the beat frequency $(\lambda _{xy}+\lambda _{z})\Omega
/\theta =(\epsilon _{1}-\epsilon _{2})\Omega /\theta $. This complex
amplitude is precisely the square-root of our resonant-normal form (\ref%
{normalform}), which is a constant because it is the modulus of the complex
number $uZ=\sqrt{C}\exp [i(\epsilon _{1}-\epsilon _{2})\Omega t/\theta ]$.
Translating the $u$ mode to Cartesian coordinates with Eq. (\ref{gyroscopic}%
), we obtain the following approximation to $B_{rad}^{(1)}$

\begin{equation}
B_{rad}^{(1)}\propto \frac{2uZ}{r}\exp (i\Omega t),  \label{second_orderB}
\end{equation}%
and therefore the frequency of the emission line is equal to $\Omega $ plus
the frequency of $uZ$ 
\begin{equation}
w_{DF}=\Omega +(\epsilon _{1}-\epsilon _{2})\Omega /\theta ,
\label{emission}
\end{equation}%
with $\Omega $ given by Eq. (\ref{orbital}). Notice that the emitted
frequency of the DF model is naturally different from the orbital frequency.
Among all possible settings of electrodynamics, only DF has this
possibility. The fact that the emission frequency of hydrogen is different
from the orbital frequency is another famous conundrum for the usual \ DR
electromagnetic setting. The emission frequency of Eq. (\ref{emission} )
contains differences of eigenvalues of the linear infinite-dimensional
operator (\ref{ZXY}) and is strikingly similar to the Rydberg-Ritz
combinatorial principle of quantum mechanics for the emission lines.

To discuss the width of the emission lines, we notice that for small
amplitudes the resonant normal form constrains the deformation while
allowing a radiative recoil. Even though the first term of the resonant
normal form is a function of the relative separation only, a higher-order
term might not be so, and at some point the recoil breaks the metastable
orbit down. The construction of this next term shall be left for future
work. In the dynamical process of resonant dissipation, the sharp line is
emitted while the dynamics is locked in the neighborhood of the original
orbit. We conjecture that when the metastable orbit breaks down, it can fall
into the next metastable attractor; another circular orbit, or into the
ground state. The stiff modes used in Eq. (\ref{pair}) describe a fast
(stiff) time-scale of a frequency of the order of $\sigma /\theta \simeq 1400
$ times the orbital frequency, the time for a stiff jump of the dynamics.
After this fast timescale the resonance essentially locks the dynamics to
the neighborhood of the metastable circular orbit.

The present simple approximation does not yet compete with the precision of
QED, but it is surprising that recognizing the correct qualitative dynamical
behavior of the equations of motion of classical electrodynamics can take us
so far (the small corrections to Eq. (\ref{ZXY}) will be given elsewhere).
The results are also beyond the Darwin approximation to Fokker's Lagrangian
because the logarithm is not analytic at $\theta =0$. We see then that the
physics of resonant dissipation is qualitatively beyond the non-ionization
criterion previously used by us\cite{discrete}, even though it contains it.
Our stability condition forces the particles to stay bound during
dissipation, a bound state condition analogous to the normalizability of the
wave functions of Schroedinger's equation.\ \ Our stability analysis
naturally involved an infinite-dimensional linear eigenvalue problem, Eq. (%
\ref{ZXY}), which is essentially a PDE, like Schroedinger's equation. We
conjecture here that it should be possible to construct an average linear
operator from Eq. (\ref{ZXY}) with $g=1$ and Eq. (\ref{ZXY}) with $g=-7$
such that $\Omega +(\epsilon _{1}-\epsilon _{2})\Omega /\theta $ is a
difference between its consecutive eigenvalues. That would be the
Schroedinger's eigenvalue problem. Last, the angular momenta of Eq. (\ref%
{hbar}) depend only logarithmically on the mass ratio times $\theta ^{4}$,
such that the deuterium and the muonium atoms have essentially the same
values for $\theta $ (i.e., the same fine structure constant, in agreement
with QED). Notice that if the proton mass is set infinite, the quantized
angular momenta become infinite as well, such that this logarithmic $\sigma $
is a genuine two-body effect.

Recognizing the correct qualitative dynamical behavior is an intuition that
goes a long way here; it suggested we should explore the infinite
dimensionality of the delay with stability analysis and make use of the
stiff modes. These stiff modes provide a natural integer number to label the
metastable orbits. The natural appearance of the infinite-dimensional
operator (Eq. (\ref{ZXY})), and the large body of qualitative and
quantitative agreement suggests that simple stability analysis inside the DF
electrodynamics is a consistent new version of QED; one with an $\hbar $
given by Eq. (\ref{hbar}). More work should be done to understand the
relation between those two theories.

We thank Savio B. Rodrigues for discussions and for checking the numerical
work in the complex $\lambda $ plane with MATLAB.


\begin{references}

\bibitem{StaruszkiewiczPole} Staruszkiewicz A,  1968 {\it Acta Physica Polonica} {\bf XXXIII},
1007, see eq. (15).

\bibitem{Bohr}Bohr N 1913 {\it Philos. Mag.} {\bf 26} , 1; Bohr N 1913 { \it Philos. Mag.} {\bf 26}, 476 .


\bibitem{Schild} Schild A 1963  {\it Phys. Rev.} {\bf 131}, 2762.

\bibitem{Hans} Andersen CM and Von Baeyer HC 1972 {\it Phys. Rev. D} {\bf 5%
}, 802.


\bibitem{Dirac} Dirac PAM 1938 {\it Proc. R. Soc. London, Ser. A}
{\bf 167}, 148 .

\bibitem{nointeract}Currie DG, Jordan TF and Sudarshan ECG  1963 {\it
Rev. Mod. Phys.} {\bf 35} 350, Marmo G, Mukunda N and Sudarshan ECG 1984
{\it Phys. Rev. D} {\bf 30}  2110

\bibitem{Driver} Driver RD 1979 {\it Phys. Rev. D} {\bf 19}, 1098 .

\bibitem{Grasman}J. Grasman, {\em  Asymptotic Methods for Relaxation Oscillations and Applications },
 Applied Mathematical Sciences, {\bf 63}, Springer-Verlag, New York (1987). 

\bibitem{Eliezer}Eliezer CJ 1943 {\it Proc. Cambridge Philos. Soc.} {\bf 39}, 173.

\bibitem{PRL} De Luca J 1998 {\it Phys. Rev. Lett.} {\bf 80}, 680 .

\bibitem{normalDeluca} De Luca J 1998  {\it Phys. Rev. E} {\bf 58, }, 5727.

\bibitem{discrete} De Luca J 2000 {\it Phys. Rev. E} {\bf 62}, 2060.



\bibitem{Fey-Whee} Wheeler JA and Feynman RP 1945 {\it Rev.Mod. Phys.} {\bf 17} 
157  ;  Wheeler JA and Feynman RP 1949 {\bf 21} 425 

\bibitem{Bellman}R. E. Bellman and K.L.Cooke, Differential-Difference Equations, 
 Academic Press, New York  (1963), page 393.

\bibitem{Rohrlich}F. Rohrlich, {\em Classical Charged Particles}, Addison-Wesley, New York (1965).

\bibitem{to be published}De Luca J, to be published.
\end{references}
\end{document}